# Decision-Making and Biases in Cybersecurity Capability Development: Evidence from a Simulation Game Experiment

Mohammad S. Jalali, PhD MIT Sloan School of Management, jalali@mit.edu

**Abstract**: We developed a simulation game to study the effectiveness of decision-makers in overcoming two complexities in building cybersecurity capabilities: potential delays in capability development; and uncertainties in predicting cyber incidents. Analyzing 1,479 simulation runs, we compared the performances of a group of experienced professionals with those of an inexperienced control group. Experienced subjects did not understand the mechanisms of delays any better than inexperienced subjects; however, experienced subjects were better able to learn the need for proactive decision-making through an iterative process. Both groups exhibited similar errors when dealing with the uncertainty of cyber incidents. Our findings highlight the importance of training for decision-makers with a focus on systems thinking skills, and lay the groundwork for future research on uncovering mental biases about the complexities of cybersecurity.

## 1. Introduction

### 1.1. Research motivation

The aftermaths of recent major data breaches and cyberattacks—affecting organizations from Yahoo, Target, T-Mobile, Sony Pictures, and JP Morgan to the US Democratic National Committee—reveal how critical it is for organizations to remain vigilant and act effectively in protecting against cyber incidents. In December 2016, Yahoo announced that over one billion accounts had been compromised in a recent incident [1]. In May 2017, Target agreed to pay $18.5 million in settlements and the total cost of the company's 2013 data breach reached $292 million [2]. Beyond the financial impact [3], a cyberattack may, for example, cause irreparable damage to a firm in the form of corporate liability [4], a weakened competitive position, and loss of credibility [5].

In coming years, the threat posed by cyberattacks will continue to grow as attacks become more sophisticated and organizations continue to implement innovative technologies that often— albeit inadvertently—introduce new, more subtle vulnerabilities. Research also suggests that attackers in cyberspace are not only rational and motivated by economic incentives [6], but also act strategically in identifying targets and approaches [7]—in other words, "The good guys are getting better, but the bad guys are getting badder faster" [8]. Perhaps there was a time a decade ago when cybersecurity was only a matter of "if" an organization was going to be compromised, but today it has become a question of "when," and "at what level."

Despite the proliferation of cyberattack capabilities and their potential implications, many organizations still perform poorly with respect to cybersecurity management. These companies ignore or underestimate cyber risks, or rely solely on generic off-the-shelf cybersecurity solutions. A mere 19% of chief information security officers (CISOs) are confident that their companies can effectively address a cybersecurity incident [9]. In May 2017, the WannaCry ransomware attack—a type of malware that blocks access to computer systems until a ransom is paid—affected companies worldwide, even though a patch for the exploited Windows vulnerabilities had been made available by Microsoft months earlier in March 2017 [10]. As data grows in size and value, the increase of cyber risks and escalation of privacy concerns demand that managers improve their approach to cybersecurity capability development—interventions to build such capabilities typically include improvements to technology already in place, in addition to the purchase of new technology, talent acquisition, and training, among other activities.

### 1.2. Research objective and approach

The importance of being proactive in cybersecurity capability development is well understood—it is more cost effective than taking a reactionary approach and reduces failure rates [11]. Although many executives and decision-makers are becoming aware of the significance of cybersecurity, a major question remains unanswered: Are experienced managers more proactive than inexperienced individuals in building cybersecurity capabilities? Further, can proactive decision-making be taught through an iterative learning process in a simulation environment?

In an attempt to answer these questions, we developed a management simulation game and conducted an experiment with experienced professionals in cybersecurity from diverse industries and a group of inexperienced graduate students. In our game, players decide how to invest in building cybersecurity capabilities for an anonymous company and monitor the effects of their decisions over the course of five years in the game.

We used simulations to facilitate the analysis of learning effects. Simulations are widely used for research and educational purposes, especially when experimentation is not feasible. For example, pilots are required to have a certain number of hours logged in flight simulators before flying a jet. Likewise, managers should have appropriate training before assuming leadership positions in complex organizational environments [12]. Management flight simulators are interactive tools that provide a virtual environment similar to the actual work settings of managerial decision-makers. Unlike in the real world where a bad choice may result in the failure of a business, simulations allow managers to practice decision-making skills without fear of real consequences outside of the game. Also, using a simulator allows for many more "flights" than might be feasible or economical.

Moreover, management flight simulators are potentially helpful for observing the long-term consequences of a decision or series of decisions. These simulators can also facilitate an iterative learning process—i.e., managers can implement their decisions, advance the simulation, monitor the effects of their decisions over time, reset the simulation, and iterate this process with another set of decisions. Similar applications have been developed in other fields, such as climate policy [13], strategic management [14, 15], and health policy [16, 17].

In addition to providing a tool to conduct this study, our simulation game allowed the subjects to experience and learn from the complexities of allocating resources for building cybersecurity capabilities. The game focuses on how a manager's decisions may impact his or her business, given the potential delays in building capabilities, and the unpredictability of cyberattacks.

### 1.3. Research contributions

Our study contributes to the current literature on cybersecurity in three ways: 1) It examines the effects of management experience in making proactive decisions for the development of cybersecurity capabilities, while accounting for uncertainties in cyber risks and delays in building capabilities; 2) It measures the effects of iterative learning on making proactive decisions through a simulation game tool; and 3) It applies a systems approach to improving cybersecurity using system dynamics modeling.

First, researchers have studied cybersecurity investment strategies in general (e.g., see [18-20]), in addition to the more specific question of trade-offs between proactive and reactive investment strategies (e.g., [11]). However, individuals' proactive and reactive investment decisions have received little attention—especially with regard to the uncertainties of cyber risks and misconceptions about, the lags in observing the benefits of capabilities. Our research aims to address this neglected area of study, but it should be noted that while there is a wide range of biases in individuals' decision-making, we focus only on biases with regard to delays in building cybersecurity capabilities. Second, we measure the effects of iterative learning in a simulation environment [21, 22] on proactive decision-making in cybersecurity capability development. Finally, systems thinking in general and system dynamics modeling in particular are not new approaches in the area of information science and technology (e.g., see [23, 24]), yet they are rarely applied to cybersecurity—especially to the business and management aspects of cybersecurity. Our simulation game views the problem of investment from a systematic perspective. It includes feedback delays in both building capabilities and identifying the consequences of decisions. Its underlying development uses system dynamics simulation modeling.

This paper is organized as follows: We first review the relevant literature and theoretical background information. Next, we present our research methodology, including the simulation game and the experimental setup, followed by the results, contributions to research and the literature, limitations and future research directions, and then our conclusions.

## 2. Theoretical background

Given that cybersecurity capability development is at the core of our simulation game, we first briefly review capability development. We then discuss the main theoretical framework of our simulation game that is based on the uncertainty of cyber incidents and delays in cybersecurity capability development.

From a general organizational perspective, 'capability' is the ability of an organization to produce a particular output [25]. From a resource-based perspective, Bharadwaj [26] developed the concept of IT as an organizational capability. Here, we consider cybersecurity capability as an organizational capability (similar to [27]) and focus on its development challenges within organizations—driven from managers' perceptions and understanding of the complexities of capability development.

Strategic management and organizational science literature shows that differences in configurations of organizational resources and capabilities explain much of the heterogeneity in organizational performance [28-30]; however, achieving optimal configurations of resources and capabilities is a complex task in which not all organizations succeed [31-33]. Similarly, configuration of cybersecurity capabilities is inextricably tied to organizational performance, and evidence shows that major variations in the configuration of cybersecurity resources exist from company to company [34].

As an example, consider two similar organizations: organization A and B. Organization A has already invested and allocated some of its resources to develop cybersecurity capabilities and as a result has well-defined plans for maintaining such capabilities. However, organization B does not have a similar perspective on cybersecurity and will only respond to cyber events in an unprepared, reactive mode. While these two different allocation of resources can explain differences in response speed and the effectiveness of responses in the wake of a cyberattack, the key question is what drives organization A, and not organization B, to allocate its resources to preemptive measures. In other words, what differences exist between managerial perspectives in resource allocation with respect to cybersecurity?

Effective investments in IT [35, 36], and in information security in particular [20, 37-40], have long been topics of interest in both academic and industry circles. In general, by allocating resources to cybersecurity capabilities, managers can not only effectively reduce potential losses due to cyberattacks, but also improve overall performance of their operations [20]. However, it remains that the diversion of funds away from profit-making processes and assets reduces cash flow [41]—an issue that is exacerbated in small and medium-sized enterprises where there is often little or no additional funding available for cybersecurity [42]. Furthermore, unlike investors who can diversify their holdings according to their appetite for risk, managers often have limited tenure in their organizations and have little choice in dealing with the risk that their company faces. Managers have to make trade-offs with regard to how they invest their resources to defend their systems [42], and it turns out that incentives drive managers to protect organizational assets in the short-term at the expense of planning for the long-term [41]. In our simulation game, players deal with this trade-off when they make investment decisions for building cybersecurity capabilities.

While the growing bank of literature on cybersecurity discusses the above questions to a certain extent, the answers we seek to the challenges of building cybersecurity capabilities in organizations are rooted in misconceptions about two particular aspects of complexity that have received little attention: the uncertainty surrounding cyber incidents, and delays in building cybersecurity capabilities. We discuss these two aspects next.

### 2.1. Uncertainty of cyber incidents

In conventional decision-making theories, the trade-off between risk and expected return is resolved by calculating risks and choosing among risk-return combinations [43]. A rational decision-maker invests in information security if the investment yields a positive return, or if the cost of the investment is less than that of the risk it eliminates. With these decisions, it is critical to have information not only about the likelihood of security incidents, but also about the impact of information security risks; however, when it comes to allocating resources to information security, difficulties in measuring the costs and benefits of information security investments cloud the vision of the rational decision-maker [44]. In addition, a consensus is rarely found among stakeholders regarding such cost-benefit analyses [45]. In risk management, it is difficult to measure the hypothetical impact of an event that is avoided [46]. Similarly, in the case of cybersecurity investment, it is difficult to estimate the impact of a hypothetical cyber incident. Further complications include a lack of historical data and effective metrics related to cyberattacks [47], a lack of knowledge about the type and range of uncertainties, a high level of complexity, and poor ability to predict future events [48]. Consequently, managers often make decisions based on their experience, judgment, and their best knowledge concerning the likelihood of uncertain events, such as cyber incidents [47].

While a manager's perception of risk is driven by his or her organizational and information system environment, as well as individual characteristics [49], research shows that humans in general do not have a strong intuition when it comes to low-probability, high-consequence scenarios, like cyberattacks. Intuitive assessment of probability is often based on perceptual quantities, such as distance, scale, or size [50]. Consider an example borrowed from behavioral economics: The more sharply one can see an object, the closer it appears [51], but if visibility is poor, people tend to underestimate the distance between themselves and the object.

The uncertain nature and severity of cyber threats, compounded with frequent shifts in technology acquisition and the introduction of new vulnerabilities makes it difficult for decision-makers to allocate resources for investment in cybersecurity capabilities [52]. The growing presence of cyber threats has resulted in an environment that has produced a large stream of information that focuses on technical defenses, but neglects the economics of cybersecurity investment [53]. If a company does not experience any cyberattacks—more precisely, if it does not detect any cyberattacks—there is little motivation to invest in cybersecurity. For this reason, many managers often do not envision cyber risks properly, hence, it is not surprising to observe significant gaps between managers' perceptions and the actual state of the cybersecurity of their organizations [54]. As a result, they may underestimate the frequency at which incidents could occur, and the time it takes for cybersecurity capabilities to become active in preventing, detecting, and responding to an incident.

### 2.2. Complex systems and delays in building cybersecurity capabilities

A complex system includes a web of interconnected components, among which there are potential delays. Regardless of the complexity of a system, a manager's problem solving method is often reactionary and event-oriented [55]; however, the use of event-oriented decision-making frameworks (e.g., situation and goals → problem → decisions → results) leads to a failure to understand the connections among components and potential delays between cause and effect. Delays between applying a decision and its effect create instability, increase the tendency of a system to oscillate, and push managers to attempt to reduce this perceived time gap long after proper corrective actions have been taken to restore the system to equilibrium [55]. Hence, a lack of understanding of the delays inherent in a system can lead to ineffective decision-making.

Even simple systems are not immune to this problem, as time delays and feedback loops between causes and effects can create complicated outcomes that are hard to anticipate [56]. Indeed, even highly educated and experienced individuals have been shown to perform poorly when making decisions in such conditions—see [57-62] for examples in various research settings. Despite these findings in other settings, practitioners in information security may believe that experienced managers understand the impact of delays better than inexperienced individuals, because of their extensive experience in industry.

Like other complex systems, cybersecurity systems include potential delays. For instance, the time it takes to develop and build cybersecurity capabilities is a major source of delays, often taking years. Human factors play a crucial role in cybersecurity [63], and recent analyses show that employees are often the weakest link in an organization with regard to cybersecurity [64]. As a result, organizations are urged to consider investing in cybersecurity training as a top priority [65], and to encourage protection-motivated behaviors [66, 67].

These recommendations mean nothing if organizations fail to understand the impacts of delays and fall into the trap of short-termism (in line with the productivity paradox and the lags in deriving the benefits from IT investments [68, 69]). Considering the delays required for the adoption of, or transition to new technologies, and for providing necessary cybersecurity training to employees, organizations may not see a return from cybersecurity training for several years.

In a reactive organization where managers start investing in the development of cybersecurity capabilities only after detecting an attack, the organization's computer-based information systems will not properly recover in time and will remain vulnerable. While the delay between cybersecurity decisions and their ultimate effects seems simple in theory—with which experienced managers would be familiar—the possession of such necessary intuition among managers is far from adequate.

In the next section, we discuss how our simulation game addresses the uncertainty of cyber incidents and delays in building cybersecurity capabilities.

## 3. Methods

### 3.1. Cybersecurity simulation game

Here, we describe cybersecurity capabilities and then present the simulation game and the setup of the experiment.

#### 3.1.1. Cybersecurity capabilities at the core

The National Institute of Standards and Technology (NIST) cybersecurity framework [70] includes five capability categories: identify, protect, detect, respond, and recover. For simplicity in the simulation game, we summarize these capabilities into three general categories: prevention, detection, and response/recovery capabilities. We merged 'identify' and 'protect' categories into prevention capabilities, and 'respond' and 'recover' categories into one group of response/recovery capabilities. In the simulation game, prevention capabilities help block computer-based information systems from being compromised by cyberattacks. Detection capabilities help detect systems that are at risk of, or presently under attack. Response/recovery capabilities help fix vulnerabilities and mitigate the damage done by an attack.

### 3.1.2. Simulation model in the game

The game is developed based on a system dynamics simulation model. The model includes three main entities: computer-based information systems, cybersecurity capabilities, and cyber incidents. Computer-based information systems are divided into four groups (visualized in the "Eco-system of computerized systems" section in Figure 1): 1) "systems not at risk" (systems with no known vulnerabilities), 2) "systems at risk" (systems with an unpatched vulnerability), 3) "affected systems" (an attacker has taken advantage of the vulnerability), and 4) "affected systems that are detected" (the attack has been discovered). While the categorization of computer-based information systems could be more detailed, we use the four general categories above for the simplicity of the game. We discuss them in more detail below.

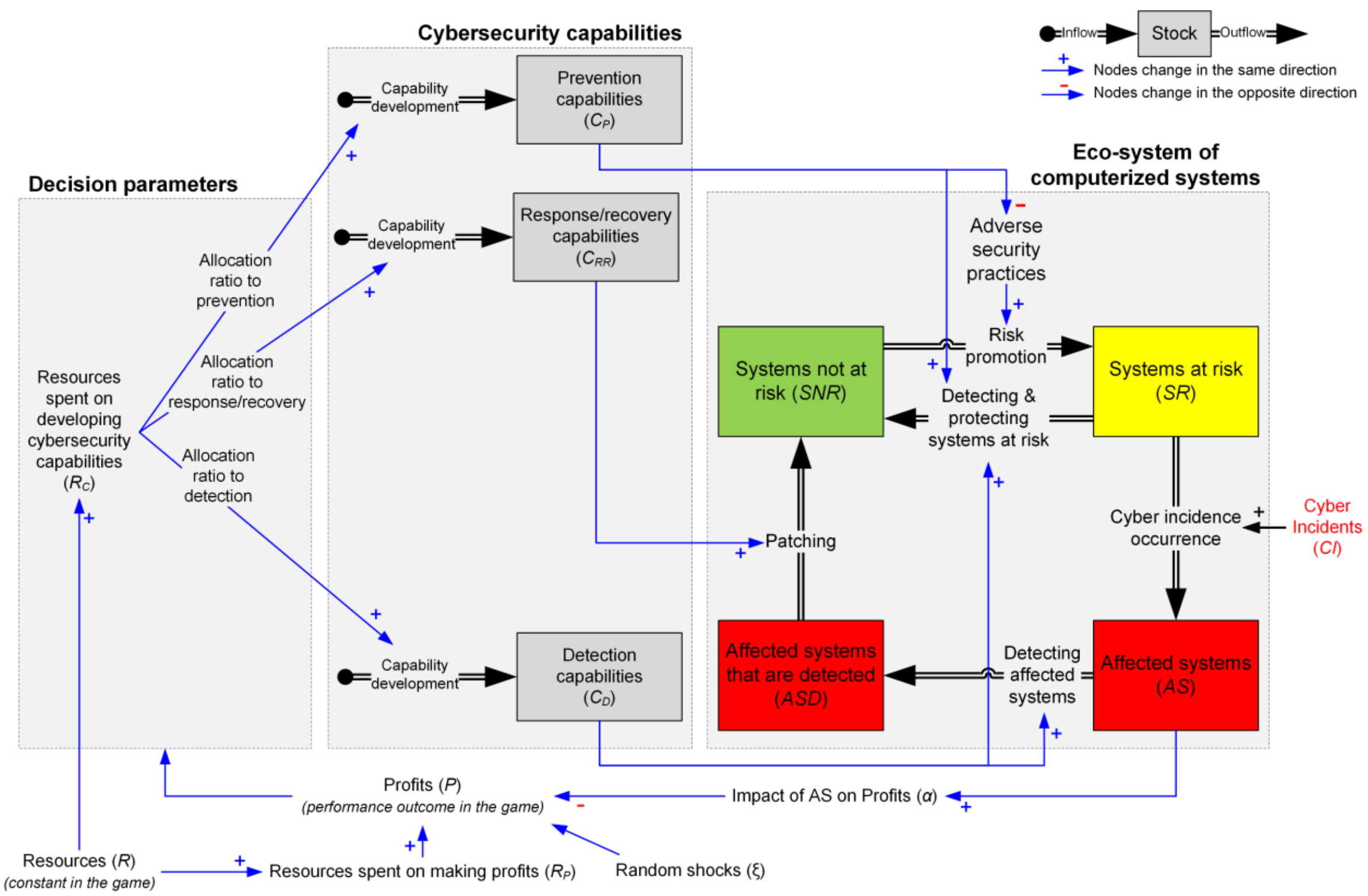

**Figure 1:** The general structure of the simulation model used in the game

### *Model structure and formulation*

Cybersecurity capabilities are divided into three groups: prevention, detection, and response capabilities. Figure 1 presents the general structure of the simulation model, with details to follow.

Players decide what percentage of their resources to spend on cybersecurity and how to distribute those resources among the three capability categories—see "Decision parameters" in Figure 1. Following the basics of organizational capability development [71], each category of capabilities ($C$) is affected by the inflow "Capability development", $I(t)$. Hence, the speed of change of the capability is $\frac{dC(t)}{dt} = I(t)$, where $t$ is simulation time. Each capability builds according to the decision parameters set by players, i.e., how much of available resources are allocated to each capability. Similar formulations are used in other settings—see [71-73].

We assume that there are multiple computer-based information systems in an organization and that all systems are initially in the "Systems not at risk" (*SNR*) group. As inadequate security practices persist, they may be moved to "Systems at Risk", *SR*. The speed at which this occurs is determined by the "Risk Promotion" rate. Prevention capabilities help protect computer-based information systems against cyberattacks and therefore decrease the flow of systems from not-at-risk to at-risk. Risk promotion is calculated as $\frac{SNR(t).ASP(t)}{b_{PR}}$, where *ASP* (adverse security practices) is calculated as $\frac{1}{C_P(t)+1}$, with $C_P$ representing prevention capabilities, and $b_{PR}$ representing the average duration to promote risk. Detection capabilities help detect cyber threats and limit the size of *SR* so that systems can be moved back to *SNR*. The flow from *SR* to *SNR* is calculated as $\frac{SR(t).(1-\frac{1}{C_D(t)+C_P(t)+1})}{b_{DSR}}$, where $C_D(t)$ is detection capabilities at time $t$, and $b_{DSR}$ is the average duration to detect *SR*.

Systems at risk are vulnerable and can be affected at the onset of a cyberattack. Once affected, a system is moved to "Affected systems", *AS*. The rate of "Cyber incident occurrence", *CIO*, is calculated as $CIO(t) = SR(t).CI(t)$, where $CI(t)$ is an exogenous fraction representing cyber incidents over the course of the simulation, which we will discuss in this section. Affected systems remain in the organization until they are detected, which depends on the adoption of detection capabilities. Once detected, systems are moved to "Affected systems that are detected", $ASD$, at a rate of $\frac{AS(t).(1-\frac{1}{C_D(t)+1})}{b_{DAS}}$, where $b_{DAS}$ is the average duration to detect that a system is in $AS$. Eventually, if proper response and recovery capabilities are in place, a system in $ASD$ can be recovered, and moved back to $SNR$. The rate at which this occurs is the patching rate, determined by $\frac{ASD(t).(1-\frac{1}{C_{RR}(t)+1})}{b_R}$, where $C_{RR}(t)$ represents an organization's response and recovery capabilities at time $t$, and $b_R$ is the average duration of the process of patching a system. $b_{PR}$ and $b_{DSR}$ are assumed to be six months and $b_{DAS}$ and $b_R$ are assumed to be two months.

It is assumed that affected systems have the potential to decrease profits. We have selected profit as the main measure of performance in the game, because monetary gains and losses are tangible and intuitive to understand. In general, the impact of cybersecurity incidents are manifested in two forms: cash flow losses and reputation damage [74]. Ultimately, both forms are translated into dollar values and we consider profits as the relevant measure which could be positive or negative. While other performance outcomes could be considered in the game, we used profit as the only measure to facilitate teasing out the effects of players' investment decisions. Also, given that organizational managers, particularly those sitting on boards or in executive roles, are typically focused on profits, it not only helps us simplify and communicate the goal of the game during experiments, but also helps players better monitor their performance and understand the effects of their decisions in the simulation.

In practice, the decision to commit to cybersecurity investment results in immediate costs associated to adoption of new technologies, including usage, and learning and switching costs, among others [75]; however, the benefits of possessing cybersecurity capabilities are harder to see than its immediate costs. Resources spent on cybersecurity investments go towards mitigating the risk of cyber incidents, which may never occur [76]. Due to the nature of cybersecurity investment, it is difficult for managers to estimate the value of cybersecurity capabilities without readily available empirical evidence.

We consider the trade-off between two major effects of allocating resources to building cybersecurity capabilities: 1) Reduced profits as building cybersecurity capabilities is an expense; 2) Possibility of protecting the organization from costly cyberattacks. Therefore, organizational resources, $R$, are either spent on making profits, $R_P$, or developing cybersecurity capabilities, $R_C$, where $R = R_P + R_C$. Hence, profits, $P$, are calculated as $P = \alpha \cdot R_P \cdot (1 + \xi)$. The impact of affected systems on profit, $\alpha$, is assumed to follow the non-linear functional form $\alpha = 1 - (AS/T)^{1/2}$, where $T$ is the total number of computer-based information systems in the organization. In our example, $T = 100$. Profit is also subject to random shock $\xi$—this exogenous shock is a pink-noise distributed normally according to $\xi \sim N(0.\sigma_\xi^2)$ [77], assuming the correlation time to be three months, and $\sigma_\xi = 0.1$.

As the players in the game can monitor profits over time and adjust their decisions accordingly, a feedback link is drawn from profits back to the decision parameters in Figure 1. The goal of the game is to maximize profits (specifically, accumulated profits over the course of a simulation run) by most efficiently allocating resources. Each simulation run covers 60 months. The trade-off between profits and protection poses a challenge, and adds to the complexity of the decision-making process already muddied by the uncertainties and delays discussed in Section 2.

It is essential to note that in the game, investments in preventive capabilities alone are not sufficient to protect the organization. Players who only invest in prevention do not have detection and response capabilities, so they are not able to detect and recover from cyber incidents. While preventive capabilities reduce the risks that an organization will be affected by cyber incidents, they can never fully eliminate those risks. Only players who strategically invest early in all three categories of capabilities are successful in reasonably protecting their organizations against cyber incidents. They observe some early reductions in total profits due to extra costs of investments in cybersecurity, but they are also able to maintain high profits over the long term.

In modeling complex systems, having a detailed and complicated model does not always translate to a more realistic simulation [78]. The simplicity of our model helps us better explain the behavior of players based on its dynamics. It also allows managers from different industry sectors to better engage with the model by relating it to their own organizations.

### *Cyber incident patterns*

Five cyber incidents with different levels of impact occur in each simulation run (the higher the impact of the attack, the more influence on profit reduction). To keep the game simple, it is assumed that all attacks are considered external attacks, in that they originate outside of the victim organization's network, and are not categorized by type, such as a phishing attack or a network-traveling worm. While we acknowledge that the risk of an insider attack is significant in reality [79, 80], the simplifications we apply do not impact the relevance of our results, since most successful cyberattacks result in data loss or operational disruptions regardless of their nature. Figure 2-a shows a pattern of five cyber incidents (in Section 3.1.3, we will discuss how the pattern of the five attacks changes in the game), and Figure 2-b presents how the string of cyberattacks affects profits (compare blue and red lines).

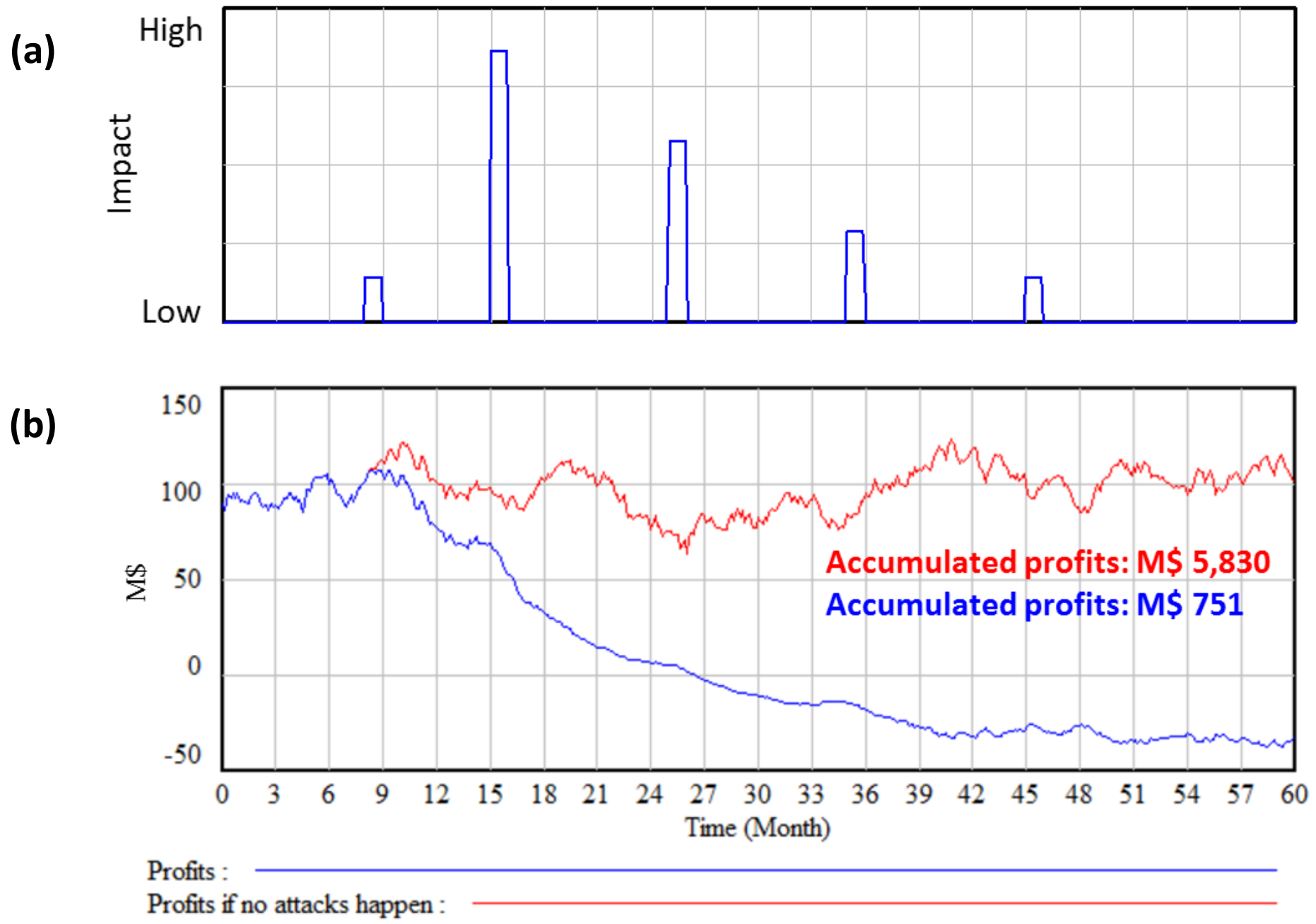

**Figure 2:** (a) One possible pattern of five attacks over the course of 60 months in the simulation, (b) Comparison of two simulations with (blue) and without (red) the impact of cyberattacks. The only difference between the two graphs in b is the occurrence of cyberattacks, and in neither simulation is any investment made in cybersecurity capability development, so any difference in profits is directly related to the cyberattacks shown in (a).

### 3.1.3. How the game works

The game runs online in an interactive environment where players have a decision parameter for each of the three types of capabilities: prevention, detection, and response. Players can adjust the value of the parameters representing the percentage of resources to allocate to each capability and when to allocate resources to each capability. Players implement their allocation strategy, advance the simulation for 12 months, monitor changes in profit and have the option to modify their allocation strategy for the next year, and advance another 12 months until 60 months, five trials, have elapsed. Each decision parameter allows the player to invest 0% to 5%, an arbitrarily set range of the IT budget, in a specific cybersecurity capability. The profit graph and accumulated profits, in millions of dollars, are shown on the same screen and they are updated any time players advance the simulation. Figure 3 shows a screenshot of the online interface of the simulation game.

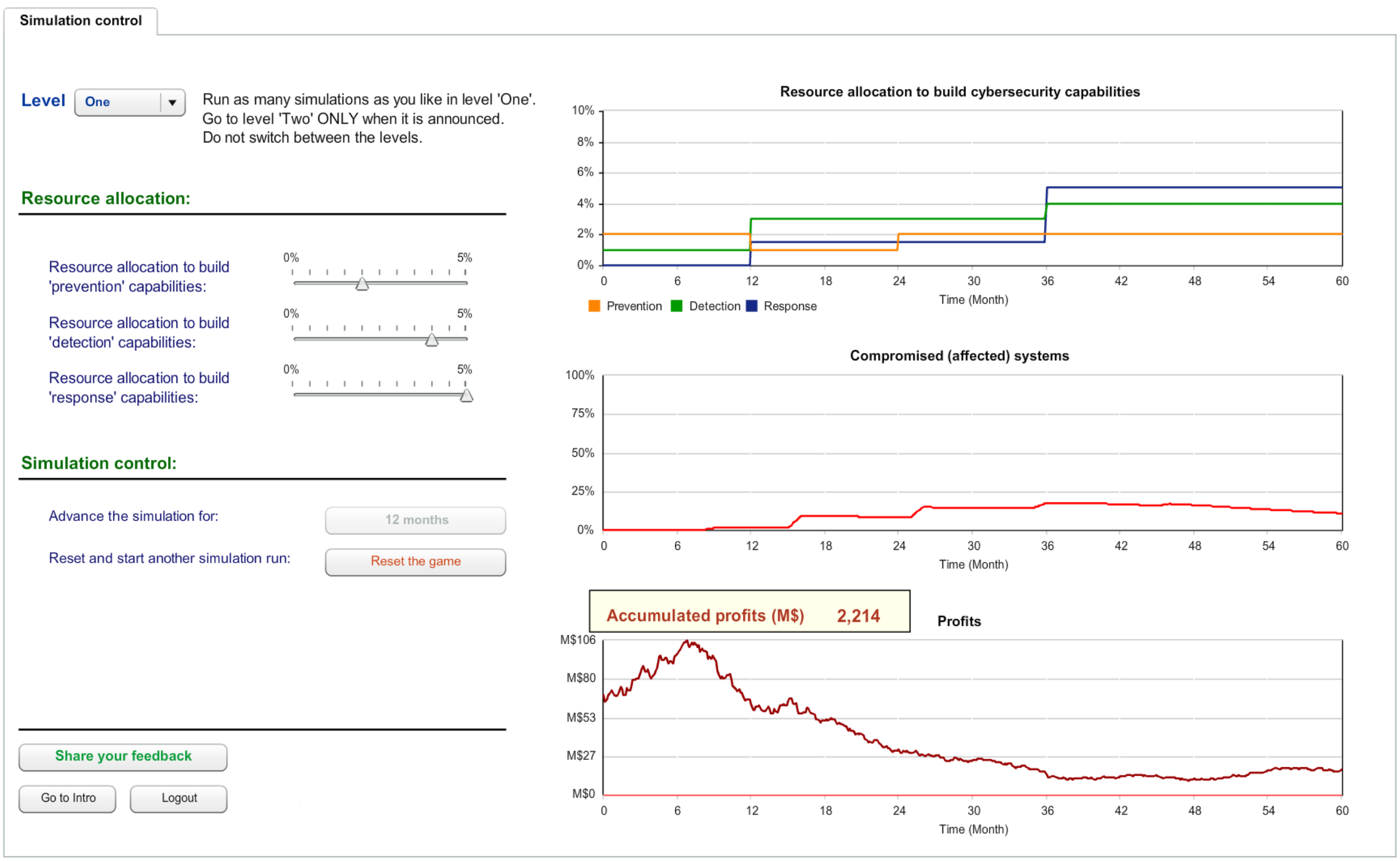

**Figure 3:** Screenshot of the online simulation game

### *Two levels of the game*

The game runs in two levels. In level one (deterministic), there are five cyberattacks with fixed impacts and at fixed times for all players. In level two (non-deterministic), five cyberattacks happen at random times with random impacts following a uniform probability distribution. All other factors remain the same between the two levels. In the interest of fairness for comparing two levels, the total impact of the five random attacks in level two is controlled to be equal for all players, and is also equal to the total impact of the five fixed attacks in level one.

Since differences in accumulated profit among players in level one can come only from differences in resource allocation, we expect that attentive players can learn how to maximize their profits after several runs; players can repeat simulations to learn about the optimal mix of investment parameters. In level two, on the other hand, unpredictability of attacks means that players face greater uncertainty when making decisions.

### 3.2. Experimental setup

We aimed to compare the performance of an experienced group to that of an inexperienced group in both a deterministic and non-deterministic setting. We discuss the setup of the experiment below.

#### 3.3.1 Subjects

Players were divided into two groups: the experienced (experimental) group, and the inexperienced (control) group. The experiment group included participants at a cybersecurity conference in Cambridge, Massachusetts, and totaled 38 professionals, with an average of 15 years of experience in IT and cybersecurity in a variety of industries. The control group included 29 Master's degree candidates enrolled in a general course about information technology. The experiment was conducted at the beginning of the semester so that any class materials would have minimal effect on the performance of the students, none of whom had had any prior experience in IT or cybersecurity.

#### 3.3.2 Experiment methods

The two groups of players were tested separately. In each case, subjects played the simulation game in the same room at the same time. To avoid social influences [81], players were not allowed to talk to each other or reveal their results at any time during the game. They were given necessary background information prior to the start of the experiment in a short presentation about the three available capabilities and watched an example play-through. They were also told that the goal of the game was to maximize accumulated profit after a 5-year period and that the person with the highest profit would be awarded at the end of the game. It was made clear that no cybersecurity capabilities were already in place at the beginning of the game. Prior to playing the game, players also had the opportunity to run the simulation themselves in a practice mode and could clarify any questions they had about the game.

Subjects first played level one for ten minutes, then played level two for another ten minutes. During each ten-minute session, subjects could run the simulation as many times as they wanted. Players were not made aware of the difference between the two levels until the results of the experiment were presented at the end of the game.

## 4. Results

In each individual simulation run, we collected data of the three investment decision variables (prevention, detection, and response) as well as profit data over the duration of each run (60 months). We also collected accumulated profits at the end of each full run to measure individuals' performance. In the experiment and control groups, 14% and 10% of the runs were incomplete, respectively (i.e., the player left the simulation run before getting to the last month), and were thus excluded from the analysis. Table 1 presents the summary of the data included in our analysis.

**Table 1:** Data summary

| Group | Number of players | Experience in IT or cybersecurity (years) | Number of individual runs | | Median of the number of runs per player | |
|---|---|---|---|---|---|---|
| | | | Level one | Level two | Level one | Level two |
| Experienced | 38 | 15 | 431 | 361 | 9 | 8.5 |
| Inexperienced | 29 | 0 | 342 | 345 | 12 | 12 |

In the following sections, we first review two individual runs as examples of proactive and reactive runs. We then present a comparison between the results of the control group and the experiment group and highlight key findings.

### 4.1. Proactive vs reactive

Figure 4 shows a proactive run and Figure 5 shows a reactive run. The trade-off discussed in Section 3.1.2 can be seen in these two figures. When proactive players start investing in capability development they make noticeably less profit than the reactive players in the early stages of the game; however, once cyberattacks begin to occur, proactive players perform much better over the rest of the simulation.

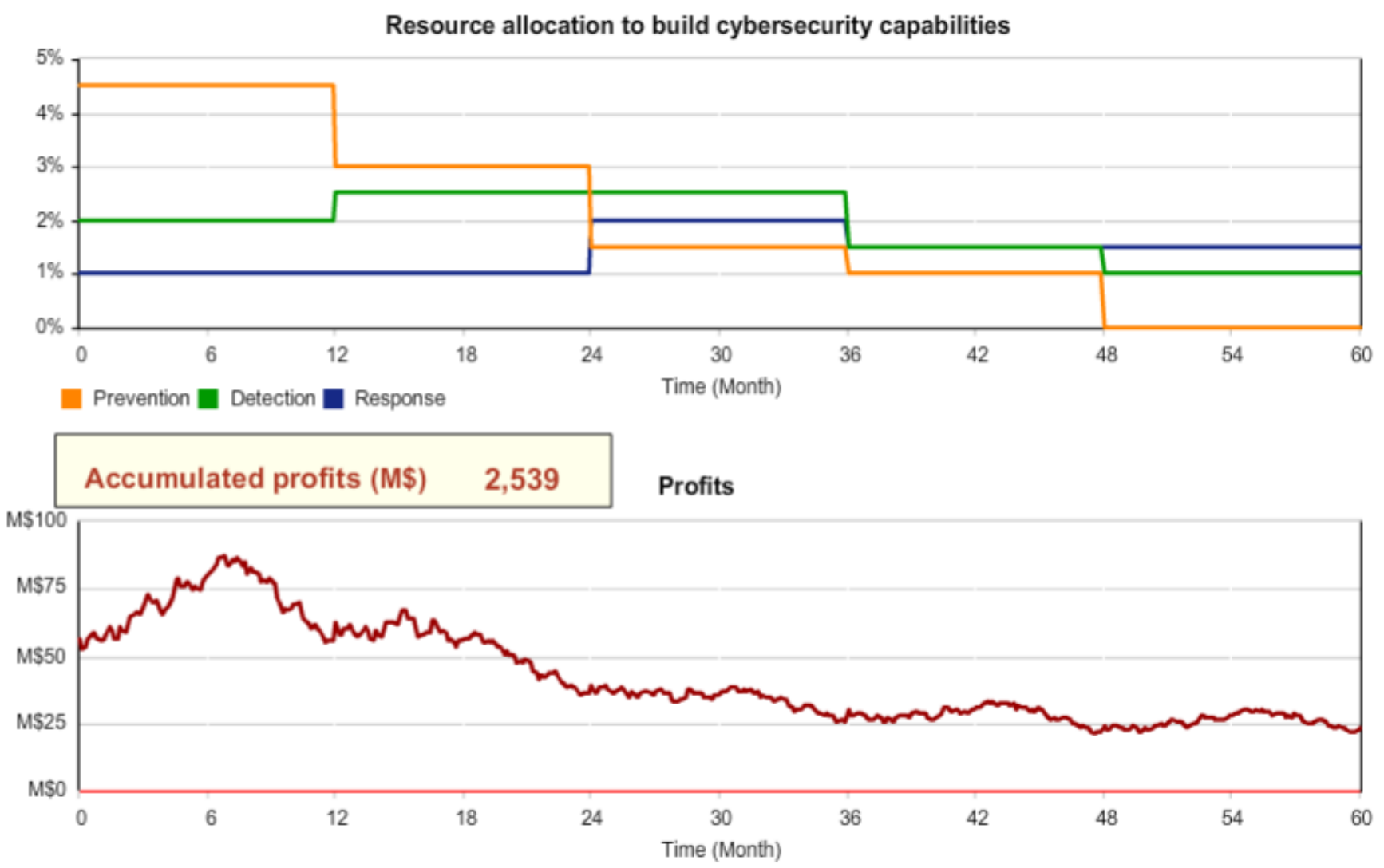

**Figure 4:** An example of a proactive simulation run

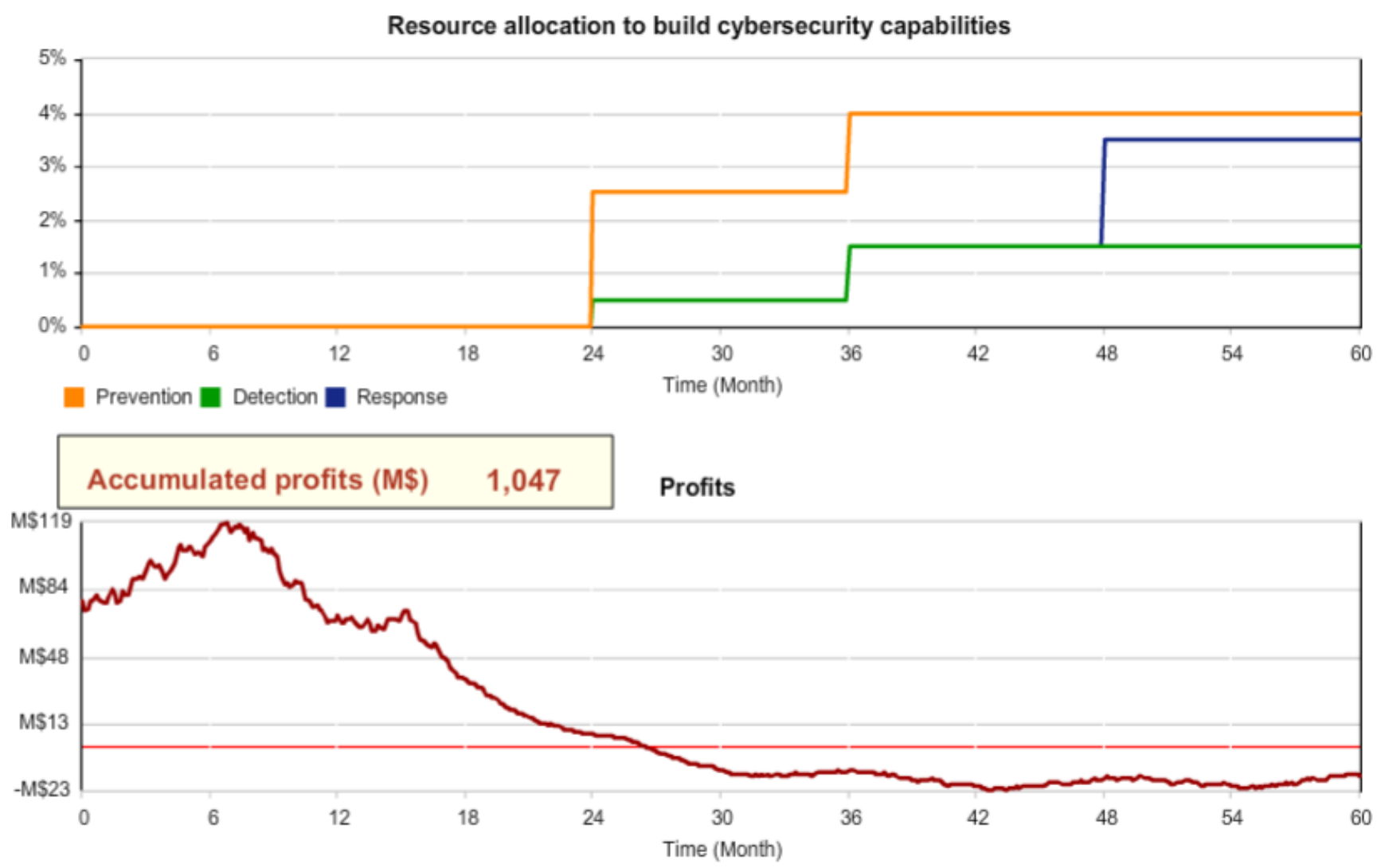

**Figure 5:** An example of a reactive simulation run

## 4.2. Comparison of the group results

Figure 6 shows the normalized distributions of subjects' best performances for the two levels of the game. Both experienced and inexperienced groups displayed large variations in performance in levels one and two; however, this variation is almost twice as large in level two as it is in level one. Given that the pattern of the five cyberattacks is fixed in level one, there exists an optimal set of values of the three decision parameters to maximize accumulated profits. Consequently, the distribution is skewed left for the poor performances, while some players are clustered close to a profit of M$2,500 on the right extreme of the distribution, indicating the maximum accumulated profit. In level two, the variability in performance increases since attacks occur randomly. Figure 6-b shows the distribution of performances in level two. It should be noted that the maximum accumulated profits in level two can be higher or lower than those of level one, due to the randomness of cyberattack occurrence.

We analyzed whether experienced players performed better in level one than inexperienced players, because only proactive investment leads to better performance in level one. One might expect experienced players to perform better than inexperienced players; however, comparing the means of the distributions using the t-test reveals that no difference exists at the 5% significance level. In other words, players with managerial experience in cybersecurity do not operate more proactively than inexperienced players.

Comparing the means of the distributions in level two (the level with random cyberattacks) using the t-test, we find that no statistically significant difference exists at the 5% significance level. Therefore, the experienced players do not make better decisions to develop cybersecurity capabilities than the inexperienced players.

Given the interesting mixed results for experienced players' performance in level one, we next analyzed the effect of iterative learning.

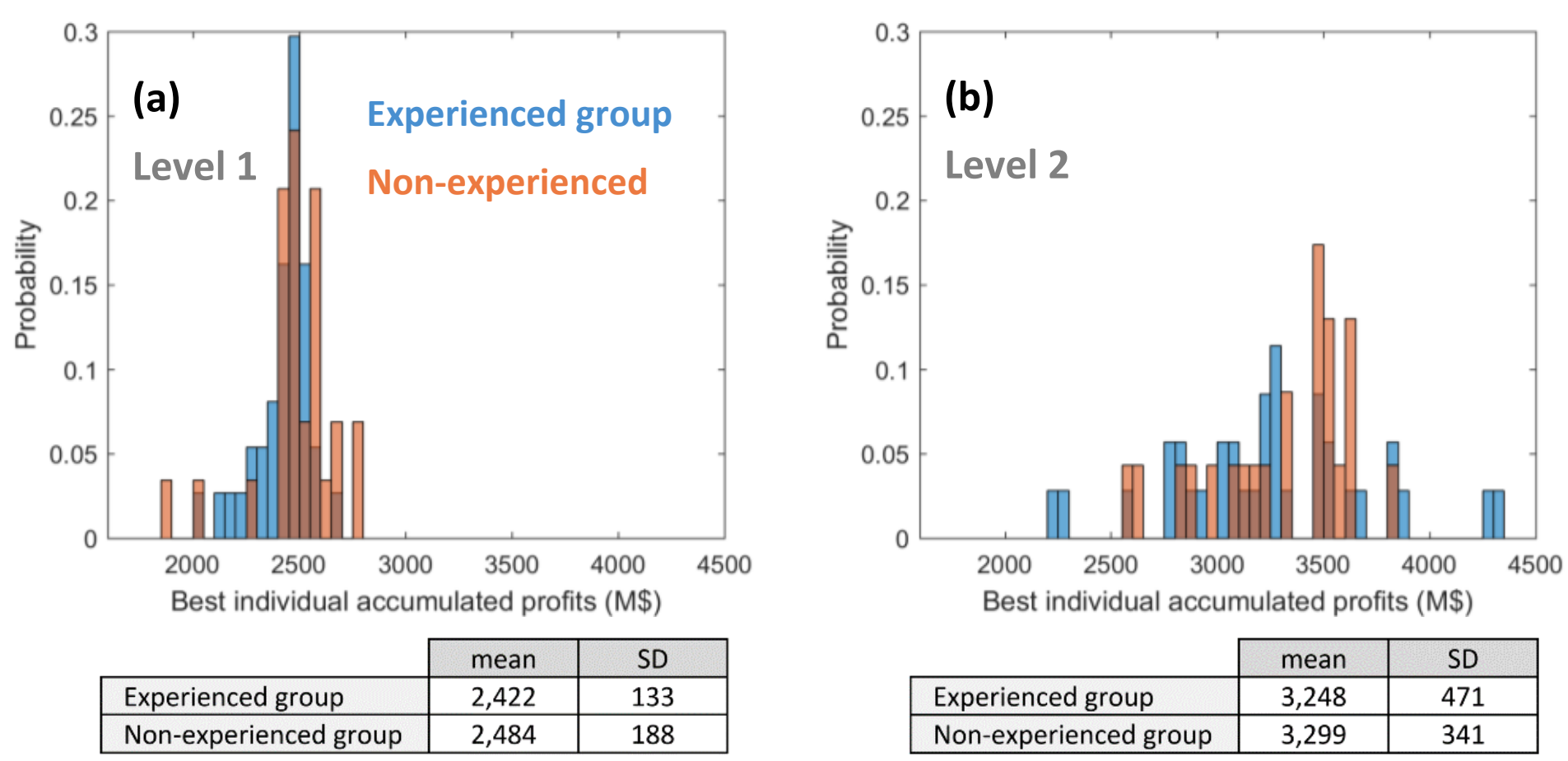

| | mean | SD |
|---|---|---|
| Experienced group | 2,422 | 133 |
| Non-experienced group | 2,484 | 188 |

| | mean | SD |
|---|---|---|
| Experienced group | 3,248 | 471 |
| Non-experienced group | 3,299 | 341 |

**Figure 6:** Distributions of best individuals' performances in level one (a) and level two (b)

### 4.3. Effect of players' iterative learning on their performance

We use correlation analysis to study the effect of iterative learning (i.e., the effect of running a sequence of simulation runs on individuals' performance). Figure 7 and Table 2 present the correlations between the number of simulation runs a player made in the 10 minutes of each level and the score of their best performances. Significantly positive correlations ($p<0.05$) are only observed for the experienced group in both levels (see Table 2). This result presents that the performance of experienced professionals significantly improved by conducting more simulation runs. As discussed earlier, high performances in the game can only be achieved by making proactive decisions; hence, the simulation game could pave the way for professionals to observe the effects of, and learn about, proactive decision-making.

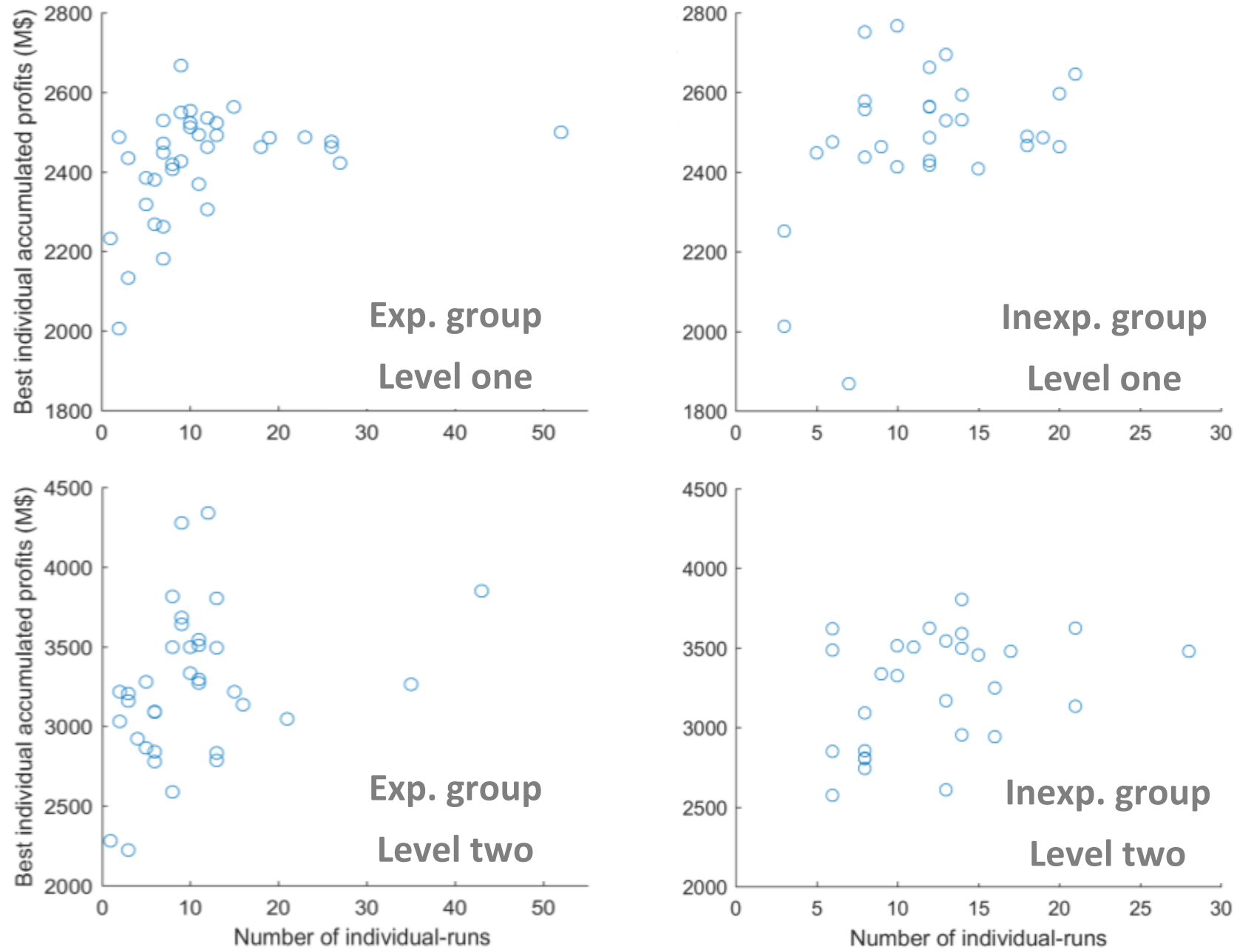

**Figure 7:** Relationships between the number of runs by an individual and that individual's best performance (each circle presents the maximum profit achieved by an individual.)

**Table 2:** Correlation between the number of runs and maximizing accumulated profits

| | Experienced | Inexperienced |
|---|---|---|
| Level one | 0.35* | 0.39 |
| Level two | 0.32* | 0.34 |

* p-value ≤ .05

Here we analyze the performance of players in each level of the game. Figure 8 and Figure 9 show the variation of players' performance (Y axis; accumulated profit) through the sequence of their runs (X axis; the number of runs for the respective player) in level one and level two, respectively.

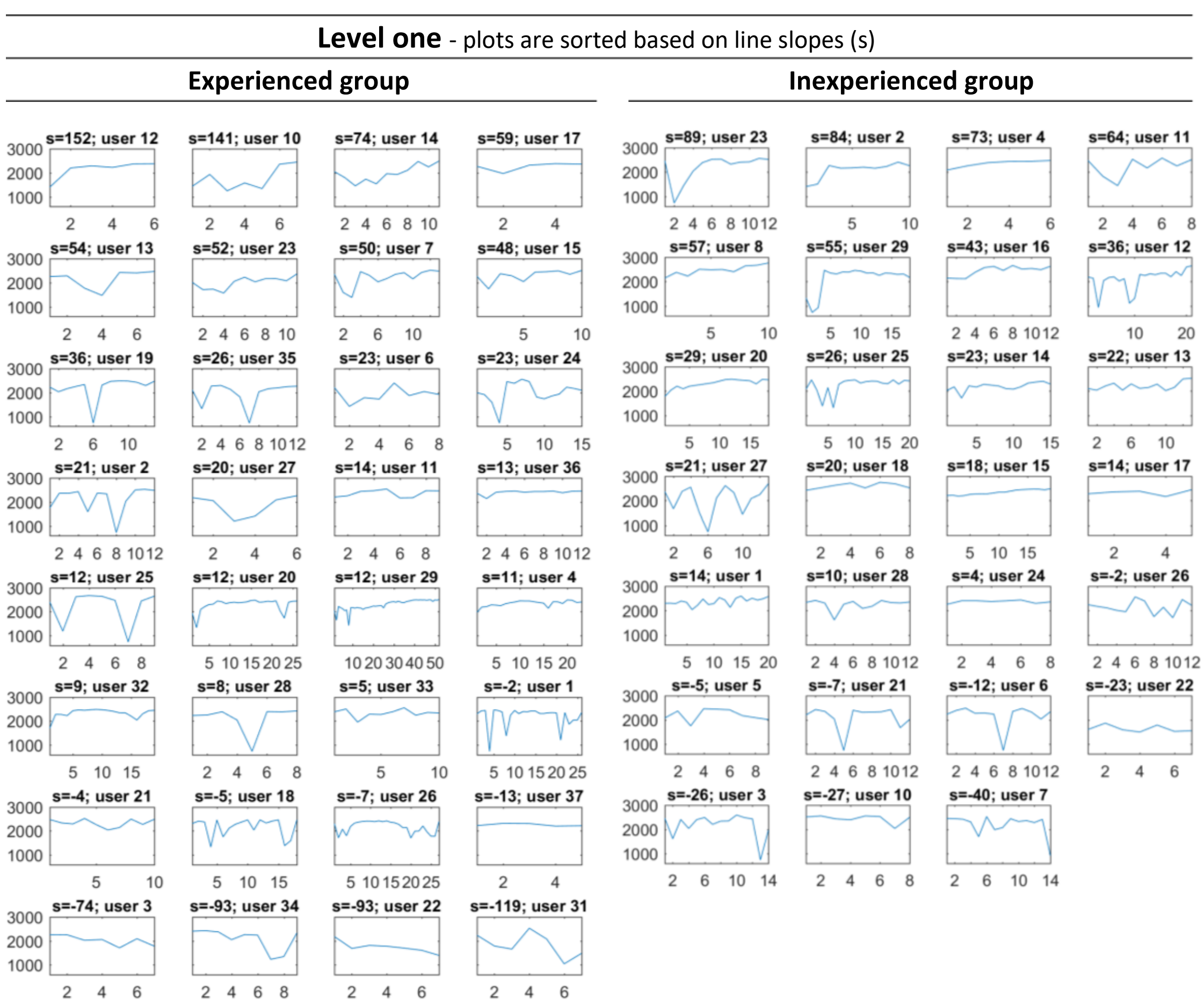

**Figure 8:** Learning curves of individuals in level one.
For each group, plots are sorted based on their linear slopes (s). The Y-axis represents accumulated profits. The X-axis represents the run number for each respective user. User IDs are unique numbers for players.

The graphs in Figure 8 and Figure 9 are sorted by linear slope in descending order, so that the graph in the top left-hand corner corresponds to the player who has shown the most improvement in accumulating profits over the course of his or her runs. For instance, in Figure 8, the top-left person in the experienced group (user 12) has the highest linear slope (s=152), which means the player improved his or her performance over the six runs in level one.

Considering the linear slope of a player's performances allows us to observe all of their run data and observe their learning process. We only considered users who played at least five rounds in a level.

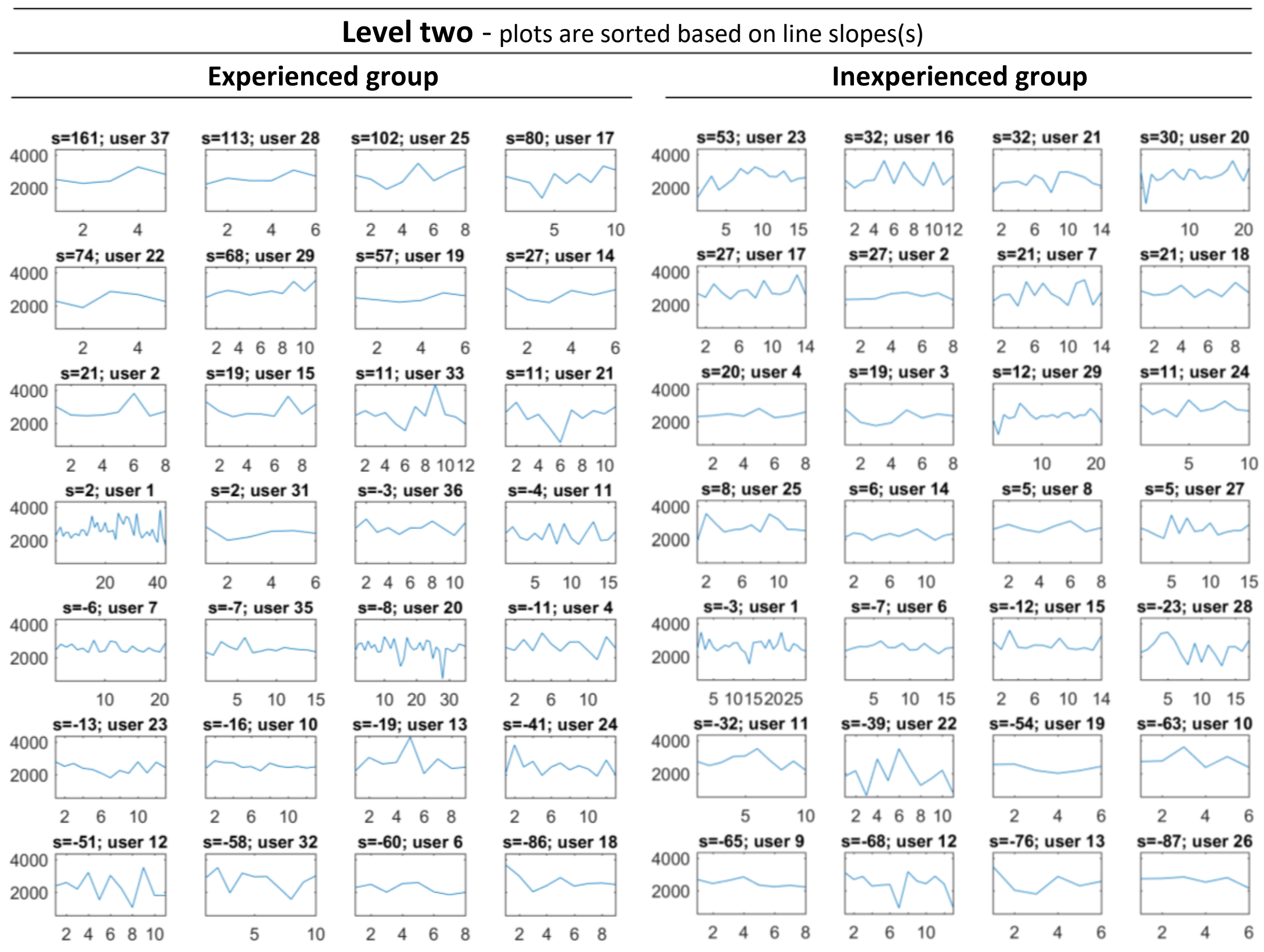

**Figure 9:** Learning curves of individuals in level two.

To interpret the results presented in Figure 8 and Figure 9, we sorted the users based on their linear slopes and analyzed how their ranks changed from level one to level two. Figure 10 presents this analysis. The line connecting the user ID from level one to level two shows the change in ranking between the levels. A red line indicates a large decrease in ranking, while a green line indicates a large rise in ranking.

Among the experienced group (the left side of Figure 10), the users who had the highest performance slope in level one performed among the worst in level two, while the users who performed poorly in level one were the better performers in level two. The top players in level one learned the game and were successful in tweaking their investment strategies to increase accumulated profits when confronted with the same fixed cyberattacks; however, the strategies they had developed did not translate effectively to the random environment of level two, and some even performed more poorly than other players who had not performed as well in level one. This pattern is not observed in the inexperienced group (the right side of Figure 10) where the ranking of players seems to change more arbitrarily[1].

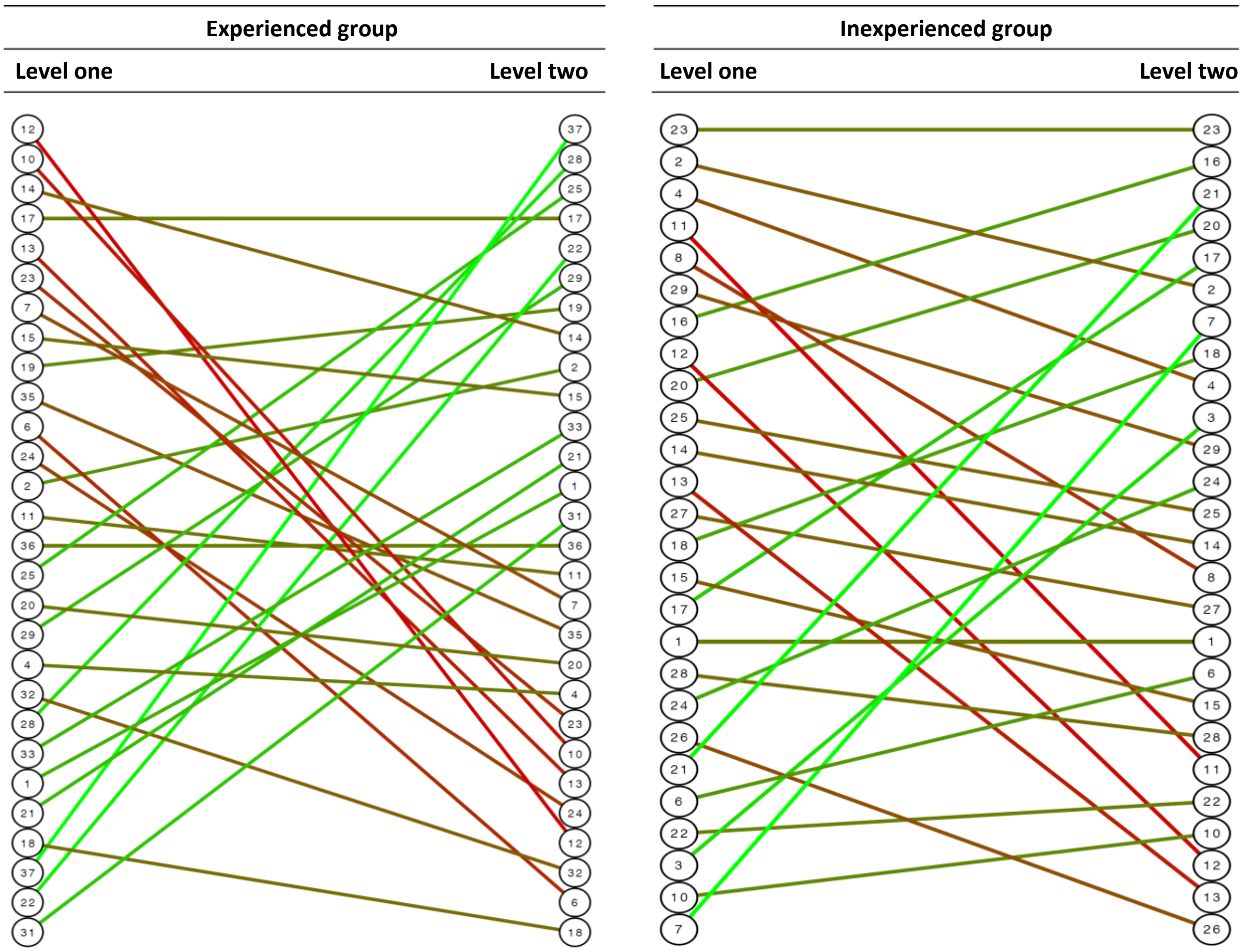

**Figure 10:** Mapping ranked users based on their learning slopes.
A red line indicates a large decrease in ranking, while a green line indicates a large rise in ranking—lighter colors represent higher changes.

[1] For instance, among the experienced group, the top six experienced individuals in level one had an average of 56% decrease in ranking in level two; however, among the inexperienced group, the top six individuals in level one had an average of 22% decrease in ranking in level two.

## 5. Discussion

We first discuss our contribution to the literature, followed by our contribution to practice. We then discuss the limitations of our research and suggestions for future research.

### 5.1. Contribution of research

Our study contributes to research on three aspects of proactive decision-making in cybersecurity: the effects of feedback delays; the role of experience; and the search for optimum decisions. We first discuss these aspects and their potential causes, as explained by prior research. We then discuss the effects of learning on making better proactive decisions.

First, our research confirms prior findings that in dynamic settings that are under uncertainty, decision-makers respond poorly to time delays (between control actions and their effects), and this strongly affects their decision performance [82-84]. If decision-makers do not compensate for feedback delays, they do not take a feedforward strategy and instead employ a feedback control strategy, which is only effective when there are no significant feedback delays [85]. In other words, they employ an open-loop approach that reduces the effectiveness of their decision performance [83, 86]. The feedforward strategy is the proactive approach needed to effectively build cybersecurity capabilities. We believe that it is reasonable to assume that problems in the performance of decision-makers are likely to be more exacerbated in the real world than the experimental setting, where they face a vastly more complex environment of cybersecurity with more feedback delays and uncertainties.

Second, in our game experiment, experienced managers did not perform better than inexperienced individuals in making proactive decisions about building cybersecurity capabilities. This is in line with findings in the decision-making literature that, in general, more experience does not equal better decision-making [87-89]. The literature indicates that: 1) Experienced individuals employ less exhaustive search procedures, utilizing processes that have worked well in the past [90]; and 2) Managers leverage their cognition for decision-making [91, 92], and their cognition could be biased by their experience [93].

Third, experienced professionals who were successful in reaching the possible optimal investment in the first level of the game were not successful in achieving near optimality in the second level. A major factor in the poor performance of experienced managers in the game is the tendency of individuals to search for alternative optimum decisions. However, as our results show, reaching optimality is hardly feasible due to the unpredictability of the times and impacts of cyber incidents. The decision-making literature suggests that there are two reasons for this poor performance: individuals' reliance on satisfactory results; and their limitations in optimization searches. The decision-making process for individuals does not involve any systematic attempt to optimize choice; instead, they tend to choose an action with satisfying results. Optimization requires a thorough search of all logical possibilities, but due to the brain's limited information processing capacity, it proves to be unsuccessful [93]. These limitations can make individuals prone to reasoning bias, human error, and fluctuating emotions [94]. Moreover, less experienced decision-makers in our game showed more flexibility in searching for alternatives in the second level of the game. This is consistent with prior similar findings that less experienced managers are more likely to engage in trying other alternatives, while experienced managers often tend to avoid risky actions [90].

Finally, consistent with prior research in other settings, e.g., [95, 96], our results show that the shortcomings above can be remedied through an iterative learning process using management flight simulators (i.e., simulation-based gaming environments). Managers can make and implement their decisions, advance the simulation, monitor the effects of their decisions over time, evaluate the performance of their decisions, reset the simulation, and redo this process with alternative sets of decisions.

### 5.2. Implications for practice

This research carries important implications for managing cybersecurity in organizations. First, it suggests that cognitive searches for optimizing resource allocation (i.e., individuals using their mental models and reasoning abilities to identify optimum investment decisions) for cybersecurity capability development are not the correct strategies, due to the complexities of cybersecurity capability development. We heard from several managers, through the comment box made available to players after finishing the game, that their thinking was heavily focused on searching for an optimum decision. One of the subjects said:

> *"The game demonstrates that even with a relatively simplified version of reality, determining 'optimum' cybersecurity investment is very difficult (or even impossible in the face of random attacks) - something that many senior executives still need to understand."*

Second, this research directs managers' attention to a major aspect of complexity in managing cybersecurity: feedback delays. While prior research on cybersecurity has predominately focused on risk estimation and uncertainty (other drivers of complexity), we observe that delays are often overlooked in decision-making for cybersecurity.

Therefore, there is a great need for training tools to reinforce the weakness of a sole-focus, cognition-driven approach to optimum strategy development and to better understand the impacts of feedback delays. The cybersecurity community has recently begun making efforts to intervene by involving and educating top-level managers and executives. This approach is well-supported by the research findings that: 1) Commitment by top executives is essential for adoption, implementation, and assimilation of security capabilities [97-99]; and 2) Cybersecurity concerns affect an entire organization (not just IT departments or isolated response teams) [100, 101]; hence, the responsibility for addressing such issues should belong to managers at higher organizational levels.

However, current training does not address the two training needs mentioned above. Cybersecurity training for top executives and decision-makers can benefit from our findings and include simulation game environments to challenge managers with: 1) the difficulties of cognitive searches for optimal decisions; and 2) the effects of feedback delays in decision-making. The simulation environment provides a context in which can implement various strategies in any number of repetitions without fear of real consequences. Our simulation game approach, which showed that iterative learning significantly improved proactive decision-making, can be used in developing training materials that better address the educational needs of managers.

### 5.3. Limitations and future research directions

The purpose of this study was not to focus on *how* to improve the understanding of the complexities of cybersecurity capability development, but rather to draw attention to the lack of understanding of these complexities at the managerial level. We also did not study how or in what priority various cybersecurity capabilities—such as prevention, detection, and response—should be developed. Working with experienced managers from a wide range of organizations, our observations show that many organizations that develop cybersecurity capabilities seem to take prevention and detection capabilities into consideration while ignoring response capabilities.

A possible extension to our simulation game would be to study the consequences of such an approach in the long run. On the importance of response capabilities, a cybersecurity expert, and the former White House chief information officer noted [102]:

> *"Preparing, planning, and especially testing for a cyber incident is crucial for all companies, both large and small. Whether your company has been actively managing cyber security risk for years or you are just beginning to develop an incident response capability, it is critical for boards and executives to engage employees in developing a robust, integrated approach to incident response. Unfortunately, companies too commonly put this task off and then find themselves flat-footed during a breach."*

Another limitation of this study is that we did not measure the risk tolerance held by individuals. Future research could take into account individual risk tolerance as it could help reveal insights into managerial decision-making biases with regards to cybersecurity. Future studies could also enhance our model by fitting it to empirical data of historical cyber events. Our experiment could also benefit from larger sample sizes, or the inclusion of non-security business experts as another control group. A more sophisticated version of the game could be developed, allowing players to take on different roles and play in teams, or play against other players in an 'attackers versus defenders' scenario. Interesting complexities will likely arise in a game where there are several interdependent groups, such as how attractive a particular organization is for attackers relative to others in the simulation game.

It would be interesting to study how players who have been given different initial settings perform in the simulation game. For example, a player who begins the simulation with compromised systems may behave differently from a player who begins the simulation with all systems in the 'not-at-risk' set, especially if they are informed about the situation. In another experiment, one group of players could be told to consider cybersecurity development costs as expenses, while another group is told that these same costs are investments. Further studies could also test the performance of players with various levels of targeted cybersecurity training. Experiments may also introduce different vectors of cyberattacks and different types of defenses in which players could invest. Moreover, studies could consider various time horizons for simulation runs (e.g., random duration of each game instead of the fixed five-year duration in our experiment) to reduce possible end-game strategizing.

Future studies could also focus on the source of biases in decision-making. Behavioral economics research shows that managers rely on a set of heuristics to make decisions in situations with uncertainty, and that these heuristics can potentially cause systematic errors. This phenomenon has received little attention in cybersecurity research, which is concerning, as a manager's biased assessment of cyber risks could hamper an organization's ability to respond to cyberattacks.

Two major heuristics, availability and representativeness, could be considered in the development of simulation games in the future. Availability refers to how readily examples come to an individual's mind [103]. People assess the likelihood of risks by considering past experiences. For example, if accidents or natural catastrophes have not occurred to them in the past, people are less likely to buy insurance. Similarly, if managers have not experienced—or are not aware of—any major cyberattacks that have affected their own organizations, they are less likely to take cyber risks seriously. Thus, the probability of cyber events occurring, given that cyber events may not be visible, is estimated to be lower than it is in actuality. The second heuristic is representativeness [103]. Representativeness describes how a certain event that shares similar characteristics with another set of events is often grouped together with the other events. The fact that something is more representative does not make it more likely. Thus, individuals relying too heavily on representativeness to make judgments are liable to make a decision for the wrong reasons [104]. Representativeness can result in serious misconceptions, especially in situations involving randomness and sequences of random events. If a person flips a fair coin five times and gets head every time, they may find it difficult to believe that the coin is indeed fair. However, if they flip the coin many times (e.g., 100 times or more), they will observe that the sequence is truly random [103]. Making decisions about uncertain cyber events is subject to this same type of error. Managers may believe that cyber events are not random and will likely look for patterns informed by past events to confront uncertainty. Using availability and representativeness heuristics together increases the chance of making the wrong decision in an uncertain world and future studies could study their impacts on decision-making in cybersecurity capability development.

## 6. Conclusions

Using our simulation game tool, we have focused on understanding how managers make proactive investment decisions for building cybersecurity capabilities. In general, there are two properties that contribute to the difficulty of making informed, proactive decisions:

- The trade-off between allocating resources to profitable activities versus investing in cybersecurity capabilities, with the perceived payoff for the latter being affected by the 'delay' between their development and reaching full functionality.
- The uncertainty surrounding the occurrence of cyber events, as shown in the differences in results from the two levels of the simulation game.

Our experiment results present that neither experienced nor inexperienced players showed performance differences in the game. This suggests deeply entrenched decision-making heuristics, reinforced over years of experience, and is supported directionally by our analysis of learning curves in the experienced group. Experienced players who performed well in level one did not perform well in level two, and vice versa, suggesting that their strategies did not adapt to the environment in which they found themselves. Inexperienced players, on the other hand, showed much more dynamism in adapting to repeated runs and the changing environment between levels one and two (see Figure 10). This is not to say that we do not recommend hiring experienced managers, but we conclude that management experience alone does not help when making decisions related to cybersecurity. Our results showed significantly positive learning effects on proactive decision-making among experienced managers. Therefore, training for better understanding of the complexities of cybersecurity is essential for experienced managers to improve their decision-making, and management flight simulators proved to be effective training tools.

Overall, our findings highlight the importance of correctly training decision-makers about cybersecurity capability development. We hope that our findings motivate the cybersecurity community to design and adopt enhanced educational and training programs that challenge entrenched mindsets and encourage proactive cybersecurity capability development. We note that the main mechanisms of our game are based on general capability development dynamics not limited to cybersecurity. The insight from this study of the importance of proactive decision-making can be applied in other settings as well, but it remains that the organizational aspect of cybersecurity, and particularly, decision-making for the development of cybersecurity capabilities, has not received adequate attention.